\documentclass[journal=jaclcd,manuscript=article]{achemso}
\usepackage[version=3]{mhchem} 
\usepackage[T1]{fontenc}       
\usepackage{amsmath}
\usepackage{graphicx}
\usepackage{color}
\usepackage[normalem]{ulem}

\author{Benhui Yang}
\affiliation
{Department of Physics and Astronomy and Center for Simulational Physics, University of Georgia, Athens, GA 30602, United States}
\email{byang@uga.edu}
\author{Chen Qu}
\affiliation{Department of Chemistry, Emory University, Atlanta, GA 30322, United States}

\author{J. M. Bowman}
\affiliation{Department of Chemistry, Emory University, Atlanta, GA 30322, United States}
\author{Dongzheng Yang}
\affiliation{Department of Chemistry and Chemical Biology, Center for Computational Chemistry, University of New Mexico, Albuquerque, NM 87131, United States}
\author{Hua Guo}
\affiliation{Department of Chemistry and Chemical Biology, Center for Computational Chemistry, University of New Mexico,Albuquerque, NM 87131, United States}

\author{N. Balakrishnan}
\affiliation{Department of Chemistry and Biochemistry, University of Nevada, Las Vegas, NV 89154, United States}
\author{R. C. Forrey}
\affiliation{Department of Physics, Penn State University,
   Berks Campus, Reading, PA 19610, United States}
\author{P. C. Stancil}
\affiliation
{Department of Physics and Astronomy and Center for Simulational Physics, University of Georgia, Athens, GA 30602, United States}
\email{pstancil@uga.edu}
\title{ Inelastic Triatom-Atom Quantum Close-Coupling Dynamics in Full Dimensionality:  all rovibrational mode quenching of water due to H impact on a six-dimensional potential energy surface}

\begin{document}

\begin{tocentry}
\includegraphics[width=6cm, height=4.5cm]{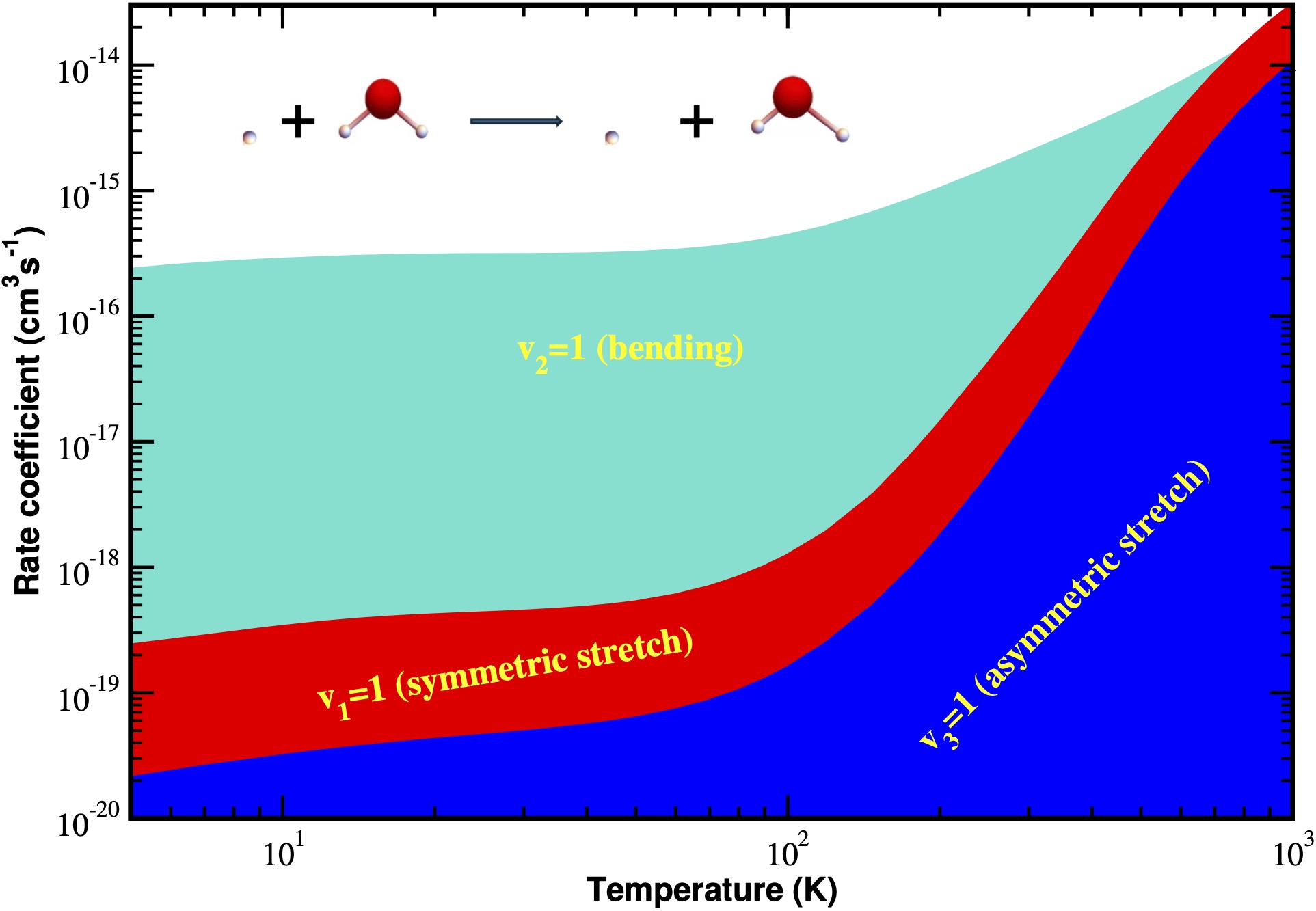}
\end{tocentry}

\begin{abstract}

The rovibrational level populations, and subsequent emission in various astrophysical environments, is driven by inelastic collision processes. The available rovibrational rate coefficients for water have been calculated using a number of approximations. We present a numerically exact calculation for the rovibrational quenching for all water vibrational modes due to collisions with atomic hydrogen. The scattering theory implements a quantum close-coupling (CC) method on a high level ab initio six-dimensional (6D) potential energy surface (PES). Total rovibrational quenching cross sections for excited bending levels were compared with earlier results on a 4D PES with the rigid-bender close-coupling (RBCC) approximation. General agreement between 6D-CC and 4D-RBCC calculations are found, but differences are evident including the energy and amplitude of low-energy orbiting resonances. Quenching cross sections from the symmetric and asymmetric stretch modes are provided for the first time.  The current 6D-CC calculation provides accurate inelastic data needed for astrophysical modeling.
 
 \end{abstract}


Water is one of the most abundant and ubiquitous species in warm regions of the interstellar medium (ISM) with a rich spectrum of rotational and rovibrational transitions. 
Since the detection of H$_2$O maser $6_{16}$-$5_{23}$ rotational transition 
near 22 GHz in 1969 \cite{chu69}, water has been  widely observed in a variety of 
diverse astrophysical environments \cite{wri00,car04,nic15,bau18,dav17,asa23}.
Water was the focus of a key program for the  Herschel
Space Observatory, Water In Star-forming regions with Herschel (WISH)  \cite{van21}. 
The  Mid-Infrared Instrument (MIRI) on-board the James Webb Space Telescope (JWST) can probe hot and warm water in shocks and inner disk surface layers
through the 6 $\mu m$ vibration-rotation band and pure rotational lines for $\lambda >10~ \mu$m.  The ground-based telescopes VLT-CRIRES+ and Keck-NIRSPEC, and in the future ELT-METIS, can also spectrally resolve
H$_2$O rovibrational emission lines at 3 $\mu m$ in star- and planet-forming disks at radii of $\sim$100 AU and constrain their location
through systematic velocity moment maps. Moreover, ELT-METIS can spatially resolve the lines down to a few AU
 and distinguish a disk surface layer from a disk wind. 
Vibrational transitions of H$_2$O were also detected using the EXES spectrometer on the Stratospheric Observatory for Far Infrared Astronomy (SOFIA) \cite{bar22,li23}.
Baudry et al. \cite{bau23} reported the observation of ten rotational transitions 
in the ground and excited
vibrational states up to ($v_1,v_2,v_3)$=(0,1,1) of H$_2$O in the ATOMIUM Band 6 survey with
Atacama Large Millimeter/submillimeter Array (ALMA).  For chemical abundance modeling, they made the following remark:
`` ... an in-depth development of H$_2$O line excitation models awaits newer collision rates and needs to
incorporate higher vibrational states, hopefully up to the (0,3,0), (1,1,0) and (0,1,1) states,
together with line overlap effects between para- and ortho-water.''
Moreover, water plays a key role in the physical and chemical evolution  and 
formation of the inner regions of protoplanetary disks (PPDs).
JWST provides opportunities to investigate the chemical properties of the warm inner regions of disks,
 and the spectral resolution of the JWST/MIRI spectrometer 
is essential to identifying H$_2$O and accurately determining its column density and temperature \cite{van23}.
Recently, Perotti et al.\cite{per23} reported JWST observations of PDS 70 and the MIRI spectrum 
showed water lines in the 7 $\mu$m spectral window, 
where the rovibrational transitions of the bending mode of H$_2$O was modeled assuming a thermal population of rovibrational levels.
Further, the inner regions of the T Tauri star Sz 98 was examined using JWST/MIRI
spectra \cite{gas23}. Both  rovibrational and pure rotational emission of H$_2$O were detected  in the emitting layers.
 
The water molecule is a typical asymmetric top with the three vibrational 
modes of H$_2$O designated $(v_1, v_2, v_3)$. $v_i$ represent the vibrational quanta of normal modes for
symmetric stretch, bend, and asymmetric stretch, respectively, with the vibrational ground state designated as $(0,0,0)$.
The rotational states of H$_2$O are labeled by $j_{K_a,K_c}$, where $j$ is the total rotational angular
momentum, $K_a$ and $K_c$ are the projections of $j$ on the molecule-fixed $a$
and $c$ axes of H$_2$O, respectively.
Water has two nuclear-spin isomers, para ($I=0$, singlet) and ortho ($I=1$, triplet),
determined by the overall spin of the two hydrogen nuclei ($I_{\rm H}=1/2$).
Para-H$_2$O exists in rovibrational states with even $K_a+K_c+v_3$ and ortho-H$_2$O exists with odd $K_a+K_c+v_3$ \cite{ten01}.

For the non-reactive collision of the water molecule with atomic hydrogen, 
a number of potential energy surfaces (PESs)  
\cite{zha91,dag13,mcc21,cab22} have been developed.
Zhang et al. \cite{zha91} reported the first three-dimensional (3D) ab initio potential for the rigid water-atomic hydrogen system
based on Moller-Plesset fourth-order perturbation theory and 
Dagdigian and Alexander \cite{dag13} developed another 3D PES 
using the restricted coupled-cluster single, double, and perturbative triple (RCCSD(T)) method. H$_2$O was
treated as a rigid-rotor, with bond length and bond angle ($r=1.8361$ bohr and $\gamma$=104.69$^{\circ}$)
corresponding to the vibrationally-averaged geometry.
A new 3D PES of H$_2$O-H was constructed by McCarver and Hinde, based on CCSD(T) calculations. In their work, H$_2$O was treated as a rigid-monomer held at its experimentally determined equilibrium 
geometry. 
Very recently, a four-dimensional (4D) PES including the H$_2$O bending mode was reported by  
Cabrera-Gonz\'{a}lez et al. \cite{cab22} The ab initio calculation was carried out using the unrestricted CCSD(T)-F12a (UCCSD(T)-F12a) 
level of theory, but with the OH bond length fixed at 0.957 \AA.

The quantum theory for scattering of a rigid asymmetric top, such as H$_2$O, with a spherical atom 
was developed in 1976 by Garrison et al. \cite{gar76} and Green \cite{gre76}.
Using the 3D PES of Dagdigian and Alexander \cite{dag13} and close-coupling formalism, 
Daniel et al. \cite{dan15} calculated rotationally inelastic rate coefficients for the first 45 levels of H$_2$O 
and for temperatures in the range of 5-1500 K.
For the rovibrational excitation of H$_2$O in collision of H,
Bissonneette and Clary \cite{bis93} performed close-coupling and coupled states approximation 
calculations of rovibrational energy transfer of H$_2$O due to H collisions. 
Cross sections were obtained for excitation of the asymmetric stretching vibration $v_3$=1 involving 
$v_1 \leftrightarrow v_3$ transitions.
A quantum study of the bending relaxation of H$_2$O due to H was presented by 
Cabrera-Gonz\'{a}lez et al. \cite{cab22} using the rigid-bender-close-coupling (RBCC) method.
The RBCC method was also used in the study of H$_2$O in collision with H$_2$ \cite{gar23}.
More recently, a full-dimensional inelastic scattering code for triatom-atom collisions, ABC+D,
was reported by  Yang et al. \cite{abcd}, based on the time-independent quantum mechanical 
coupled channel method. The ABC+D code has been successfully used to study
rovibrational scattering between H$_2$O and Cl, He, and Ar atoms. \cite{yang21,yang22a,yang22b}

 We report a new 6D, full-dimensional PES for H$_2$O-H and the first coupled-channel (CC), 
 full-dimensional scattering calculation with full angular momentum coupling for rovibrational
transitions of water due to H impact for the first excited states of all vibrational modes. The calculations 
were performed with the ABC+D code and an invariant polynomial fit of this new 6D PES. Here we briefly describe the methods used in rovibrational inelastic scattering calculations and PES calculation and fit.
 The reader is referred to Refs.\citenum{yang21} and \citenum{yang23} for additional details. 
 
The total Hamiltonian for the collision of a triatomic molecule ABC and a spherical atom in the coordinate system presented by  Brocks, van der Avoird, Sutcliffe, and
Tennyson (BAST) \cite{abcd,bast} can be expressed as (in atomic units),
\begin{equation}
\hat{H}=-\frac{1}{2\mu}\frac{\partial^2}{\partial R^2}+\frac{\mathbf{(J-j_2)}^2}{2\mu R^2}
 +\hat{h}_{\textrm{ABC}}(j_2,\mathbf{q}) + V(R,r_1,r_2,\theta_1,\theta_2,\phi),
\end{equation}
where $\mu$ is the reduced mass for the collision system and 
 $\mathbf{q}\equiv (r_1$, $r_2$, $\theta_1)$ are intramolecular Radau coordinates for the triatom ABC. 
The collision coordinate from the center of mass of ABC to the atom is described in 
Jacobi coordinates  as ($R$, $\theta_2$, $\phi$). 
 $V(R,r_1,r_2,\theta_1,\theta_2,\phi)$ is the interaction potential between ABC and atom D. 
$\mathbf{j_2}$ is the rotational angular momentum of ABC and $\mathbf{J}$ the total angular momentum. 
$\hat{h}_{\textrm{ABC}}$ is the Hamiltonian for ABC with the parity-adapted eigenfunctions of ABC
obtained by solving the equation, 
\begin{equation}
\hat{h}_{\textrm{ABC}}|\eta;JM\epsilon \rangle = E_{\eta}|\eta;JM \epsilon \rangle,
\end{equation}
where channel index $\eta$ is defined as  $\eta \equiv (j_2tK)$. 
$t$ is a rovibrational state of ABC, $E_{\eta}$  is the internal energy of the $\eta$th channel,
and $\epsilon$ is the system parity.  $K$ and $M$ are the projections of $J$ onto the body-fixed and space-fixed
$z$ axes, respectively.  The total wavefunction of the scattering system can be expanded in terms of eigenfunctions of 
the triatomic molecule:
\begin{equation}
\Psi^{JM\epsilon}=\sum_{\eta} F_{\eta}^{J\epsilon}(R)|\eta;JM\epsilon\rangle.
\end{equation}
The resulting second-order CC differential equations for the radial 
functions are expressed as \cite{man86} 
\begin{equation}
 \frac{d^2}{dR^2}F_{\eta}^{J\epsilon}(R)=\sum_{\eta'}W_{\eta',\eta}^{J\epsilon}(R)
  F_{\eta'}^{J\epsilon}(R).
\label{cceq}
\end{equation}
The coupling matrix $W_{\eta',\eta}^{J\epsilon}(R)$ is given by
\begin{equation}
W_{\eta',\eta}^{J\epsilon}(R)=2\mu V_{\eta',\eta}^{J\epsilon}(R)-\delta_{\eta'\eta}2\mu E_{c,\eta}
+ \frac{1}{R^2}U_{\eta',\eta}^{J\epsilon}(R),
\end{equation}
where $E_{c,\eta}$ is the collision energy. The centrifugal matrix $\mathbf{U}$ is expressed as 
\begin{equation} 
\begin{split}
 U_{\eta',\eta} & = \langle \eta' |\mathbf{(J-j_2)}^2| \eta \rangle  \\
 & =  \delta_{t't}\delta_{j_2^{\prime} j_2}\{\delta_{K' K}[J(J+1)+j_2(j_2+1)-2K^2] \\
 & -\delta_{K',K+1}\sqrt{1+\delta_{K0}}\lambda_{JK}^{+}\lambda_{j_2K}^{+}  \\
 & -\delta_{K',K-1}\sqrt{1+\delta_{K1}}\lambda_{JK}^{-}\lambda_{j_2K}^{-} \}, 
\end{split}
\end{equation}
where $\lambda_{ab}^{\pm}=\sqrt{a(a+1)-b(b \pm 1)}$. The matrix elements of the coupling potential, 
$\mathbf{V}$, are, 
\begin{equation}
V_{\eta',\eta}=\delta_{K',K}\langle\eta'|V|\eta\rangle.
\end{equation}

The log-derivative method \cite{man86,joh73} was applied to 
solve the CC equation (\ref{cceq}).  
Within the CC formulation, the state-to-state integral cross section for a transition from an 
initial state $j_2t$ to a final state $j_2^{\prime}t'$ is given by
\begin{equation}
\sigma_{j_2^{\prime}t' \leftarrow j_2t}  
  = \frac{1}{2j_2+1}\frac{\pi}{k_{j_2t}^2}\sum_{J}(2J+1)\sum_{K'K\epsilon}|\delta_{j_2^{\prime}j_2}\delta_{t't}
  \delta_{K'K}-S_{j_2^{\prime}t'K' \leftarrow  j_2tK}^{J\epsilon}(E_c)|^2,
\end{equation}
where the wave vector $k_{j_2t}=\sqrt{2 \mu E_c}$ and $S_{j_2^{\prime}t'K' \leftarrow j_2tK}^{J\epsilon}$ is the $S$ matrix.

The interaction energy between H and H$_2$O was computed using MOLPRO2010 \cite{molpro} at the RCCSD(T)-F12b \cite{adl07,wer07} level of theory with the aug-cc-pVQZ
basis. The basis set superposition error was removed by applying the counter-poise correction.
The data set consists of 151,027 configurations and corresponding energies. The geometries 
in the data set were sampled on a 6D grid. 
This includes 20 values for $R$ ranging from 3.5 to 30.0 bohr, 6 different values for $\theta$ over
[0, $\pi/2$], and 7 different $\phi$ values  over [0,$\pi$].

From the full data set, 75,476 points with $R \ge 8.0$ bohr were selected as the long-range data.
The final PES combines a fit to the full data set (denoted as $V_{\rm I}$) and a fit to the long-range data
(denoted as $V_{\rm II}$).  The potential energy of the combined PES, referred to as
VHH2O, is given by
\begin{equation}
V = (1-s) V_{\text{I}} + s V_{\text{II}},
\end{equation}
where $s$ is a switching function defined as,
\begin{equation}
s = \left\{
\begin{aligned}
& 0\ (R < R_i), \\
& 10 b^3 - 15 b^4 + 6 b^5 \ (R_i < R < R_f), \\
& 1\ (R > R_f),
\end{aligned}
\right.
\end{equation}
where $R_i=8.0$ $a_0$ and $R_f=10.0$ $a_0$, and $b = (R - R_i) / (R_f - R_i)$.
Both $V_{\rm I}$ and $V_{\rm II}$ have been fitted in 6D using the permutation invariant polynomial
method via monomial symmetrization \cite{zhe10}.
Both fits for $V_{\rm I}$ and $V_{\rm II}$ used a maximum polynomial order of 8, with 1589 linear coefficients.
The root-mean-square (rms) fitting error of  $V_{\rm I}$ is 2.93 cm$^{-1}$,
while the rms of $V_{\rm II}$ is 0.025 cm$^{-1}$. The PES is smooth and free of bumps or holes.

\begin{figure}[h]
\centering
\includegraphics[scale=0.6]{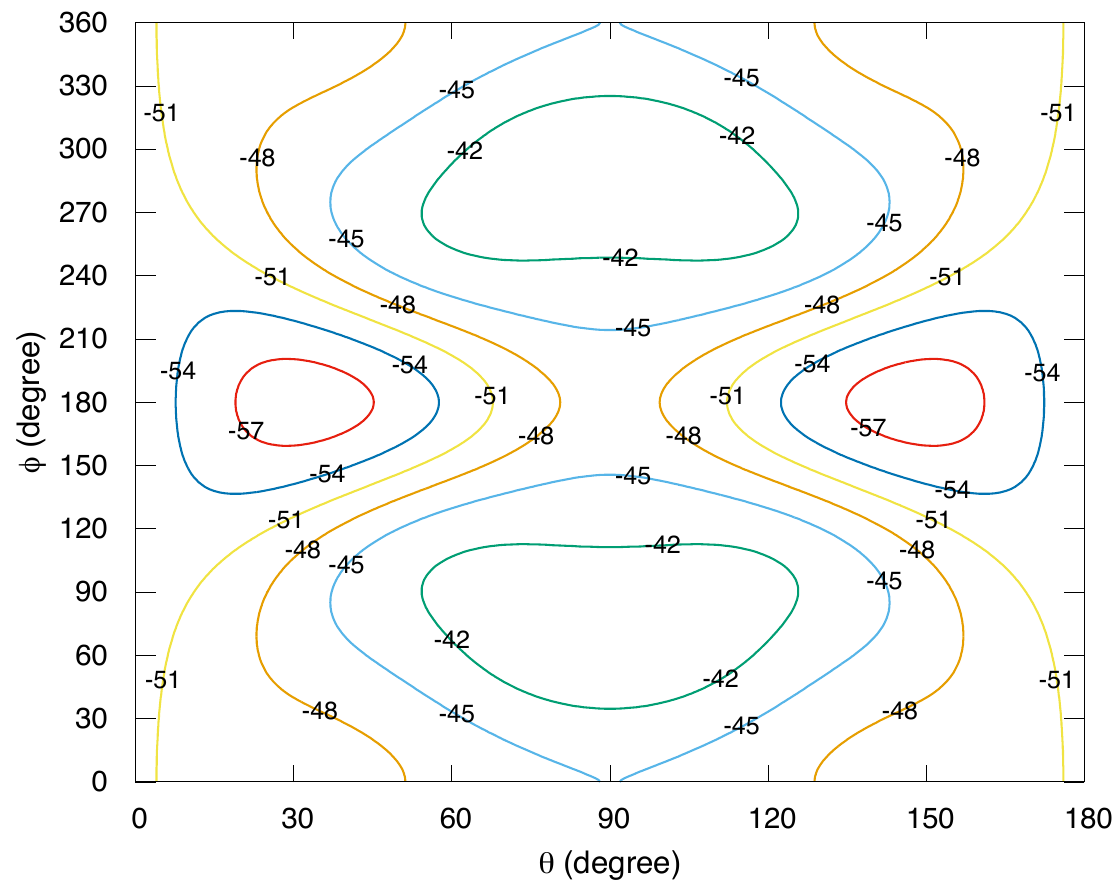} 
\caption{Contour plot of PES as a function of orientation angles $\theta$ and $\phi$, $R=6.5$ bohr. H$_2$O is treated as a rigid rotor at the vibrationally-averaged geometry.
  $R, \ \theta, \ \phi$ are body-fixed polar coordinates defined in Fig.~1 of Ref. \citenum{dag13}. The H$_2$O molecule lies in the  $xz$-plane with the origin centered on the H$_2$O center of mass with $z$ axis oriented along the A inertial axis of H$_2$O. The unit of potential energy is cm$^{-1}$.}
\label{contour}
\end{figure}

Fig.~\ref{contour} shows a two-dimensional contour plot around the global minimum of 
$\theta$ and $\phi$ for the rigid rotor H$_2$O-H potentials of the 6D PES for H-H$_2$O distance $R=6.5$ bohr. 
 Note that $R, \ \theta, \ \phi$ are body-fixed polar coordinates defined in Fig.~1 of Ref.~\citenum{dag13}, and $R$ is the 
 distance between H  and center of mass of H$_2$O.
 In Table~S1, the well depth and equilibrium  geometry of the previously reported H$_2$O-H PESs are presented. To further assess the accuracy of the 6D PES, we performed rigid-rotor inelastic scattering calculations as described in Section S2 in the
 Supporting Information. The good agreement between our rotational de-excitation rate coefficients and the results obtained with a 3D PES \cite{dan15}
 confirms the accuracy of our 6D PES.

Full-dimensional  rovibrational scattering calculations were carried out
using the ABC-D code \cite{abcd}
in which the CC equations were propagated from $R=$ 3.5 to 30.0~$a_0$.
The number of Gauss-Legendre quadrature points adopted for $\theta_1$,  $\theta_2$, and $\phi$ 
to integrate out the angular dependencies in the matrix elements  were $N_{\theta_1}$=23, $N_{\theta_2}$=19, and $N_{\phi}$=20, respectively. 
The number of potential optimized discrete variable representation (PODVR) \cite{lig00} points 
were chosen as $N_{r1}=N_{r2}=4$.
The energy truncation for the contracted basis functions employed was $E_{\eta}^{max}$=6000 cm$^{-1}$.
The log-derivative propagation was carried 
out with different step sizes $\Delta R$ in different ranges of $R$:  
$\Delta R$ = 0.04 $a_0$ in $R \in [3.5, 5.0] \ a_0$, 
$\Delta R$ = 0.08 $a_0$ in $R \in [5.0, 10.0] \ a_0$, 
and $\Delta R$ = 0.1 $a_0$ in $R \in [10.0, 30.0] \ a_0$. 
 Furthermore, convergence tests of the cross sections with respect to partial wave summation and rovibrational basis have been performed. The maximum number of partial waves $J$ in the scattering calculations is up to 100.
All these parameters yielded results converged to within 5\%.
We focus first on the vibrational
quenching from the first excited state of the water bending mode ($v_2$=1), (010).
The calculation of rovibrational state-to-state quenching  cross sections and rate coefficients was performed for initial
rotational states of para-H$_2$O: $0_{00}$, $1_{11}$, $2_{02}$, $2_{11}$, $3_{13}$, and $2_{20}$,
and of ortho-H$_2$O: $1_{01}$, $1_{10}$, $2_{12}$, $3_{03}$, and $2_{21}$. 

\begin{figure}[h]
\centering  
 \includegraphics[scale=0.4]{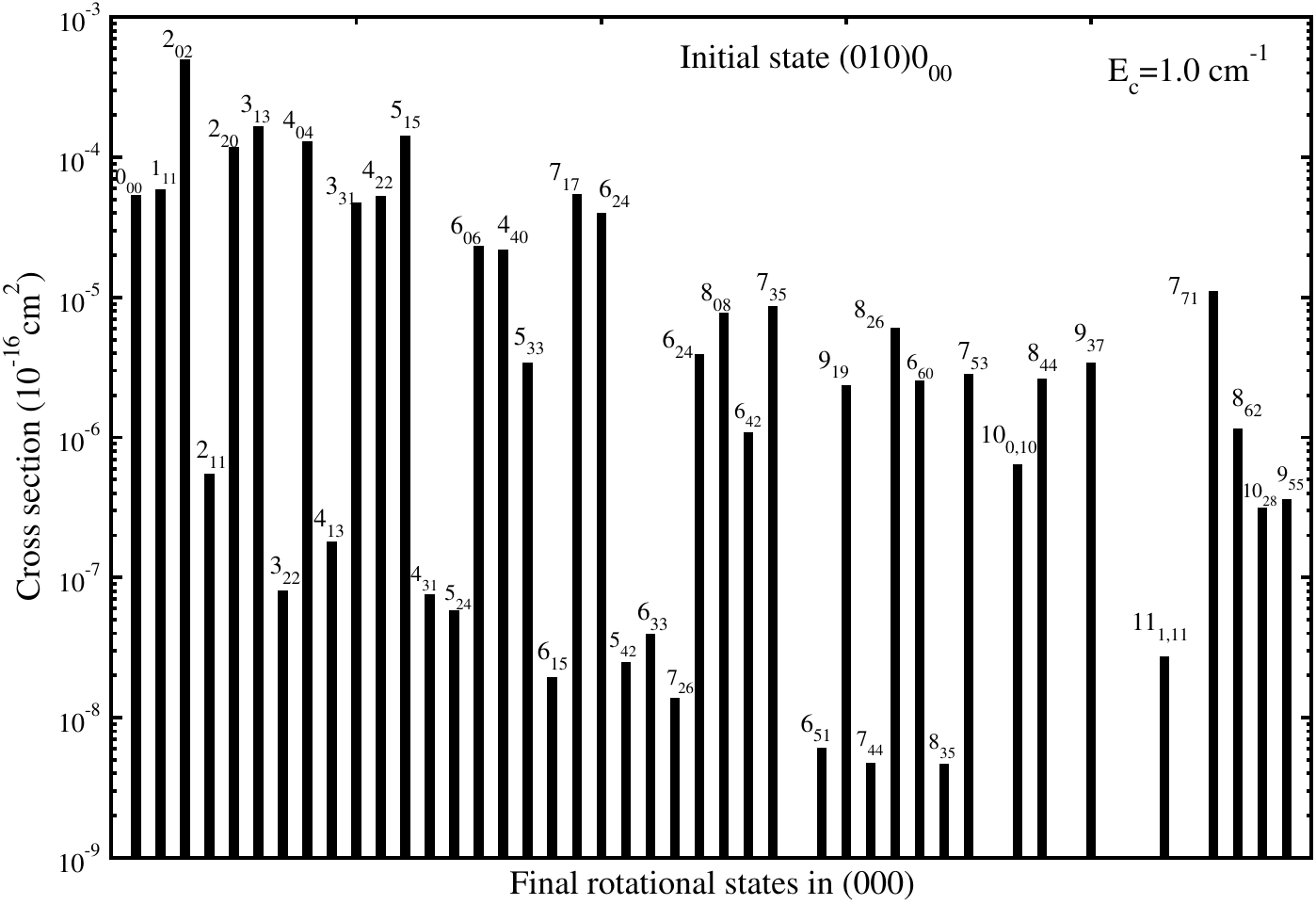}
 \includegraphics[scale=0.4]{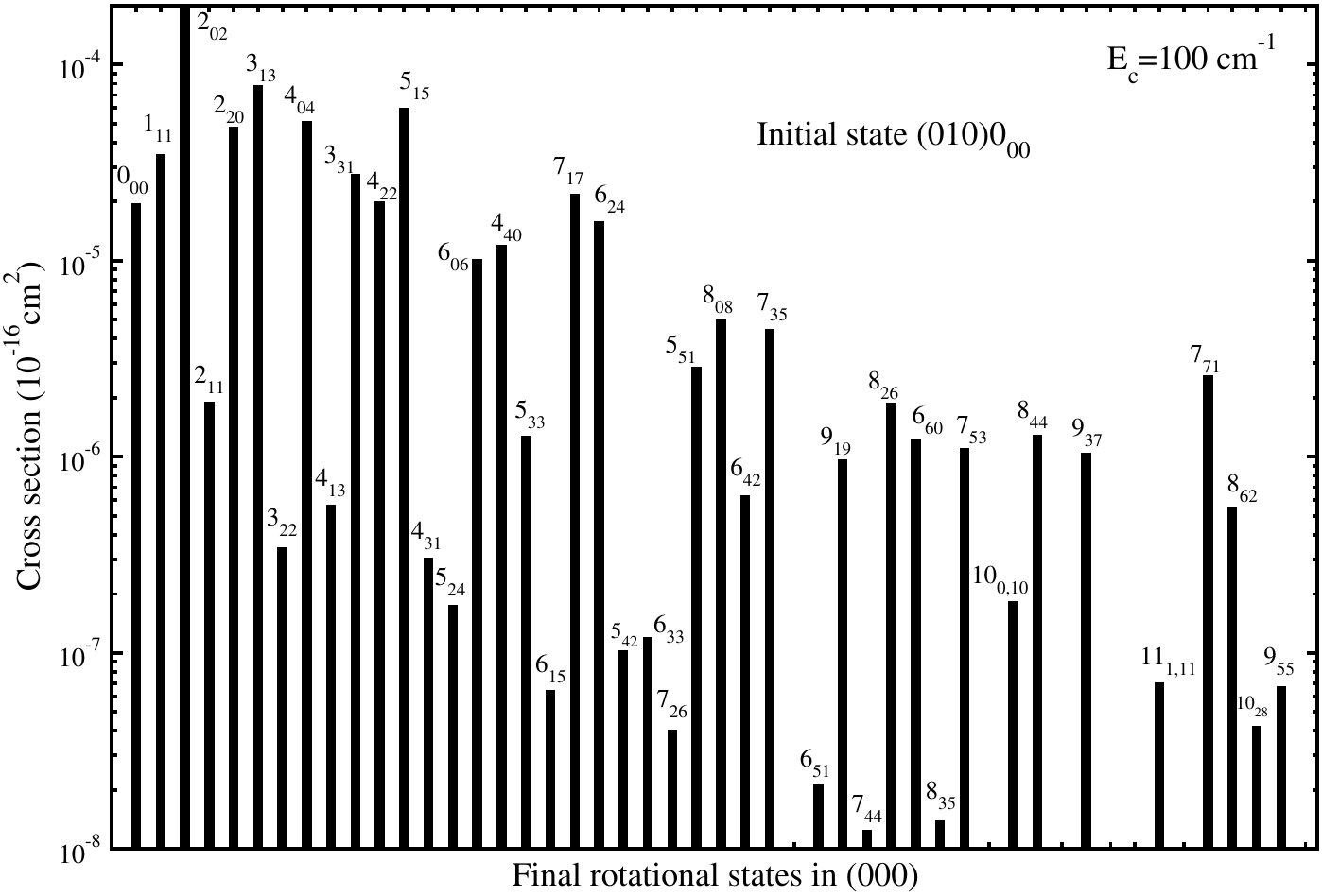}    
\caption{ 
Rotational state distribution of para-H$_2$O in (000) from initial 
state (010)$0_{00}$ in collision with H. The final rotational states are energy ordered. 
 Upper: collision energy = 1.0 cm$^{-1}$;  
 Lower: collision energy = 100.0 cm$^{-1}$.  
}
\label{rotdis_p}
\end{figure}

Fig.~\ref{rotdis_p} shows the distribution of final rotational states of para-H$_2$O in the ground vibrational state (000)  following de-excitation from initial state (010)$0_{00}$ at collision energies of 1.0 cm$^{-1}$ and 100 cm$^{-1}$.
The final rotational distribution is broad and centered in the region of low $j$ levels. 
The distributions extend up to higher rotational levels for both collision energies. No propensity rules are evident, though the dominant transition is for $\Delta j=2$, $\Delta K_a=0$, and $\Delta K_c=2$.
The weakest transitions typically have $\Delta j \neq \Delta K_a$ and/or $\Delta j \neq \Delta K_c$.
Some final states are missing in the plot because of their small cross section magnitudes.
In Fig.~S3 we also show the distribution of final rotational states but for ortho-H$_2$O in the ground vibrational state.
 The distribution patterns are similar to para-H$_2$O.

\begin{figure}[h]
\centering
\includegraphics[scale=0.85]{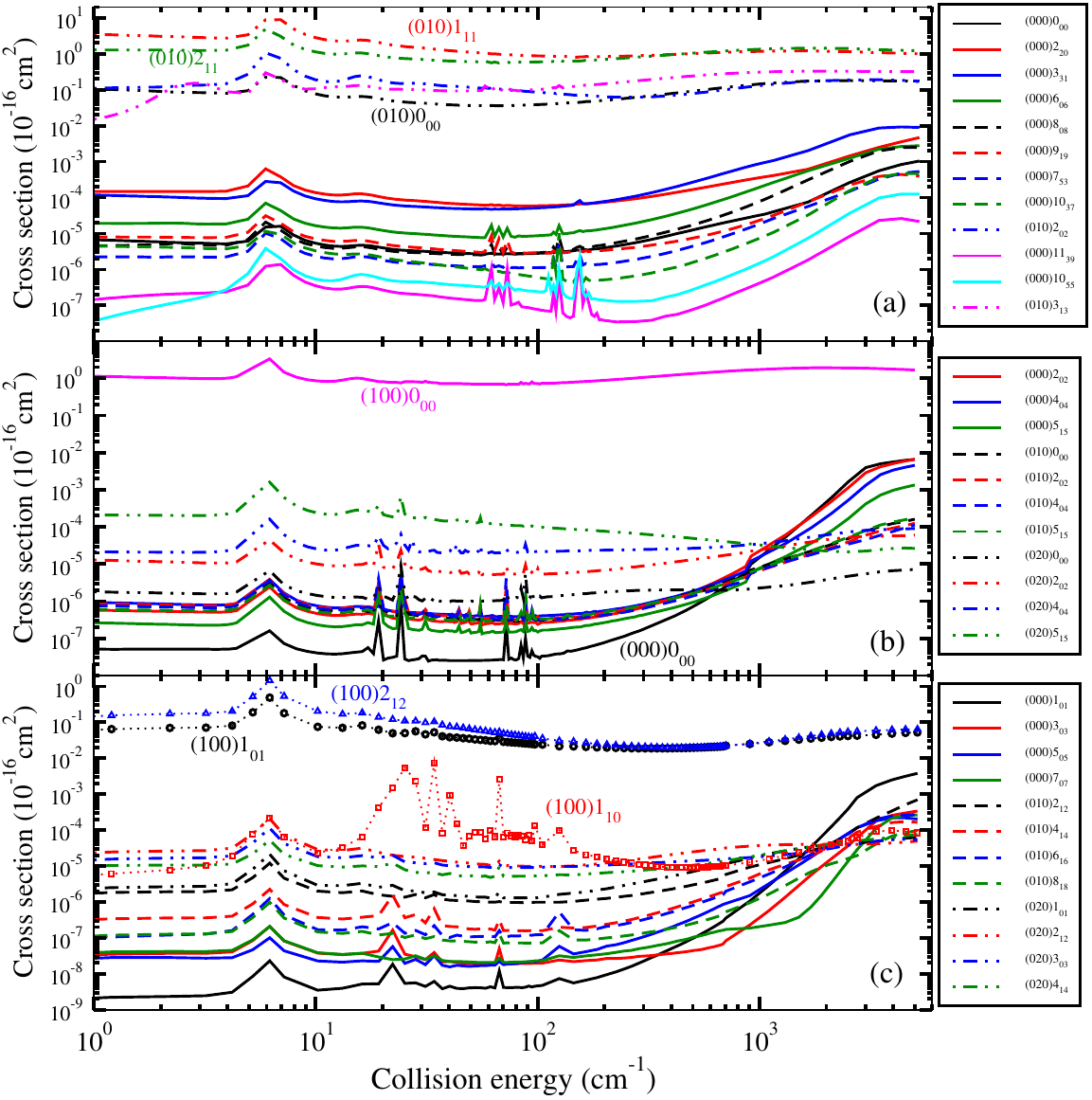} 
\caption{ 
State-to-state rovibrational quenching cross sections of H$_2$O in collision with H.
(a): from initial para-H$_2$O state (010)$2_{20}$;
(b): from initial para-H$_2$O state (100)$1_{11}$;
(c): from initial ortho-H$_2$O state (001)$0_{00}$.}
\label{s2s_cross}
\end{figure}

In Fig.~\ref{s2s_cross} we present the state-to-state cross section for rovibrational quenching from H$_2$O vibrational states (010), (100), and (001). Because of a large number of final states, as examples, we only present some selected final states for each initial state.
Fig.~\ref{s2s_cross}(a) shows the results from the initial state (010)$2_{20}$ for para-H$_2$O with H. It can be seen that two groups of cross sections are shown for $\Delta v_2=-1$  (lower) and  $\Delta v_2=0$ (upper).  Resonance structures are present for energies between $\sim$1 and $\sim$100 cm$^{-1}$. The results of ortho-H$_2$O with H shown in Fig. S4 give similar behavior as para-H$_2$O. The state-to-state rate coefficients for the vibrational quenching (010) $\rightarrow$ (000) corresponding to the cross sections shown in Fig.~\ref{s2s_cross}(a) and Fig.~S4 are presented in Fig.~S5 for temperatures ranging from 5 to 1000 K. The rate coefficients display similar trends as shown in Fig.~\ref{s2s_cross} (a) and Fig.~S4 for the cross section. 

We have also extended the full-dimensional scattering to the rovibrational quenching from the symmetric stretching vibrational state (100) of para-H$_2$O and asymmetric stretching vibration (001) of ortho-H$_2$O.  We believe that cross sections from these vibrational modes of water are computed here for the first time. The cross sections are displayed in Fig.~\ref{s2s_cross} (b) and Fig.~\ref{s2s_cross}(c) for selected final rotational states in vibrational states (000), (010), (020), and (100). As the symmetric and antisymmetric vibrational modes are higher in energy, exoergic transitions to excited bending modes (010) and (020) are accessible as 
shown in the figures.
All the vibrational quenching cross sections display resonances in the low collisional energy region due to quasibound states supported by the attractive part of the interaction potential.
To identify the partial wave contribution to the resonances at collision energies near 5 to 8 cm$^{-1}$, 
we show in Fig.~S6 the $J$-resolved contributions to the quenching cross sections to selected final states from initial state (010)$2_{20}$. The dominant partial waves are $J=$6, 5, 2, and 5 for final states (000)$0_{00}$, (000)$2_{20}$, (010)$0_{00}$, and (010)$2_{02}$, respectively.

\begin{figure}[h]
\centering
\includegraphics[scale=0.8]{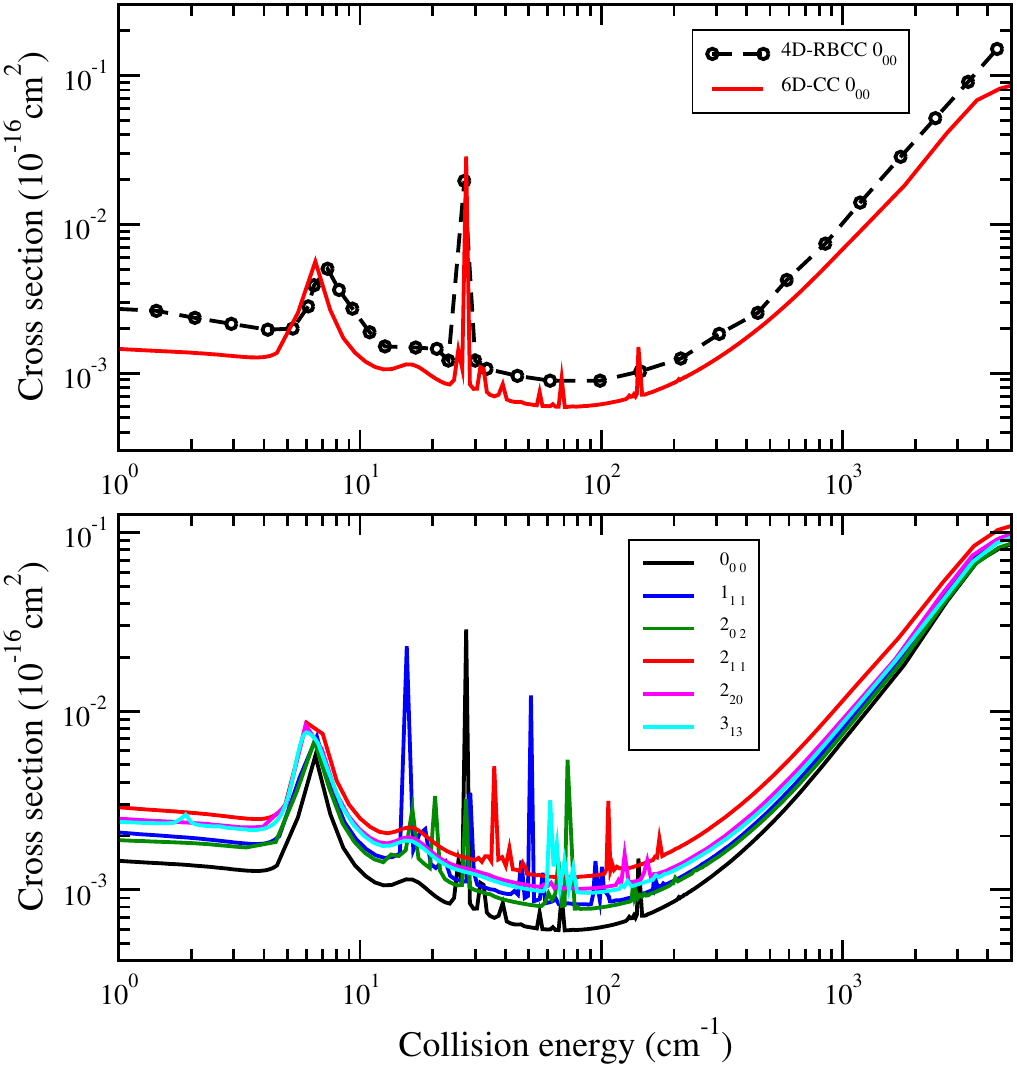} 
\caption{ 
Total quenching cross section from $(010)j_{K_a,K_c}$ to (000) of para-H$_2$O in collision  with H.
(Upper panel) Comparison between present 6D-CC calculation with the 4D-RBCC results of Cabrera-Gonz\'alez et al. \cite{cab22}
 from initial state $(010)0_{00}$.
(Lower panel) Present results from initial state $j_{Ka,Kc}=$ $0_{00}$,  $1_{11}$, $2_{02}$, $2_{11}$, $2_{20}$, and $3_{13}$.
 }
\label{ptotal}
\end{figure}

The total vibrational quenching cross section from an initial state $(010)j_{K_a K_c}$ can be obtained by summing the state-to-state cross sections over all final rotational states $j'_{K'_a K'_c}$ in (000). Fig.~\ref{ptotal} displays the total vibrational quenching cross sections of para-H$_2$O from initial states (010) $j_{K_a,K_c}=$ $0_{00}$,  $1_{11}$, $2_{02}$, $2_{11}$, $2_{20}$, and $3_{13}$. The upper panel shows the comparison between our total quenching cross section from initial state (010)$0_{00}$ with
the RBCC result of Cabrera-Gonz\'{a}lez et al. \cite{cab22}. In general, the agreement is reasonable, though the current cross section is somewhat smaller.  The two dominant resonances are reproduced in the current calculations, but additional resonances over the collision energy range of 30 to 200 cm$^{-1}$ are found due to the finer energy grid adopted here. 
The differences can be attributed to the features of the PESs used in the scattering calculations, in particular, the different well depths of the PESs as shown in Table S1. 
The lower panel of Fig.~\ref{ptotal}  shows the total vibrational (010) to (000) quenching cross section from initial states $j_{K_a,K_c}=$ $0_{00}$,  $1_{11}$, $2_{02}$, $2_{11}$, $2_{20}$, and $3_{13}$. All of the total quenching cross sections show similar behaviour and reflect the presence of a large number of resonances in the intermediate energy region between $\sim$3  and $\sim$200 cm$^{-1}$. For energies larger than 200 cm$^{-1}$,  the cross sections increase with increasing collision energy, typical of vibrational-rotational/translational energy transfer. 
Fig.~S7 displays the total quenching cross sections for ortho-H$_2$O.  Our 6D-CC calculations show good agreement with the 4D-RBCC results, but we again find additional resonances. The total quenching cross sections from initial states (010) $j_{K_a,K_c}$=$1_{01}$, $1_{10}$, $2_{12}$, $2_{21}$, and $3_{03}$  are also shown in Fig.~S7.

\begin{figure}[h]
\centering
\includegraphics[scale=0.65]{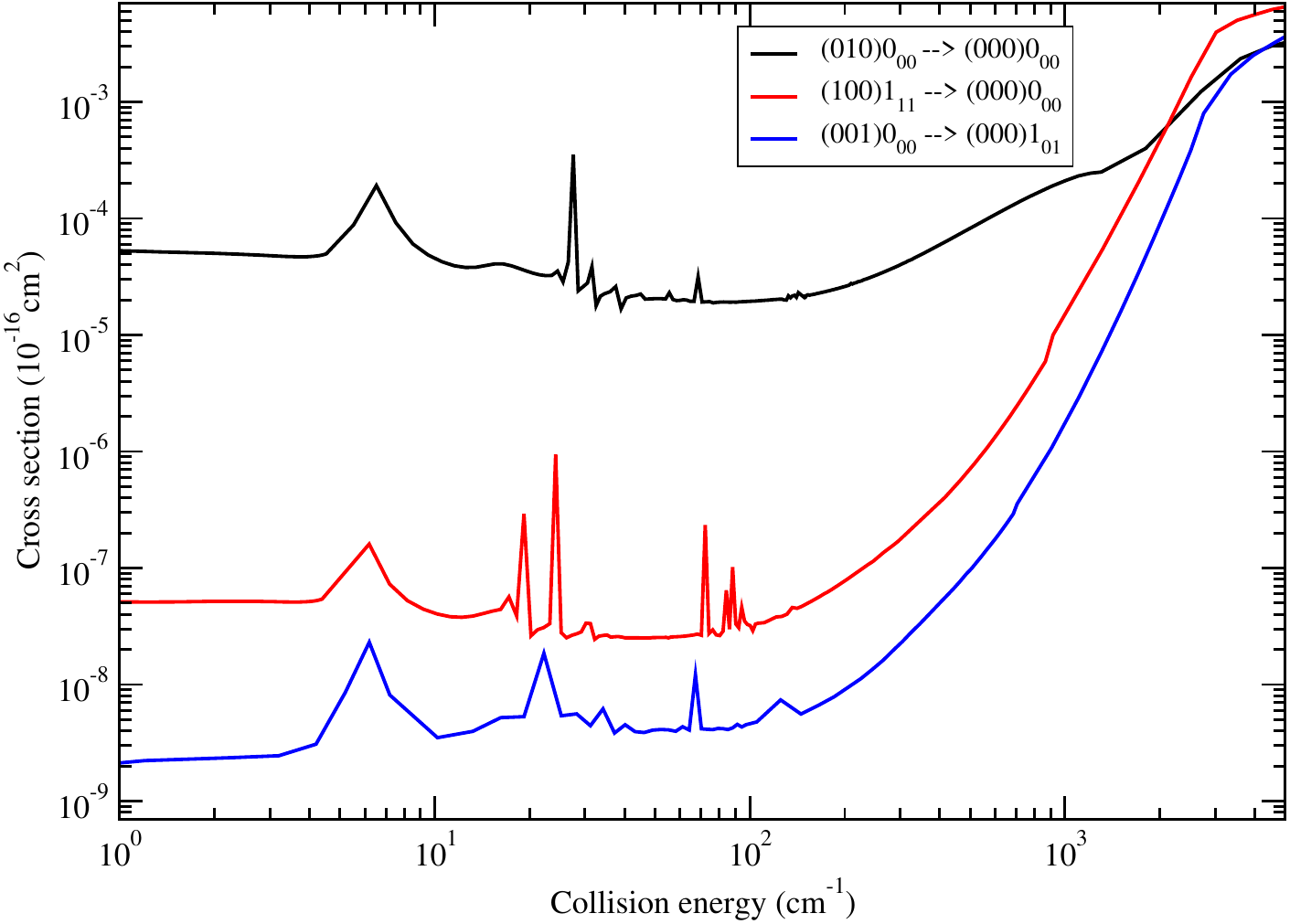} 
\caption{ 
State-to-state quenching cross section from  initial states $(010)0_{00}$, 
$(100)1_{11}$, $(001)0_{00}$ to the lowest rotational states in the ground vibrational state  (000) of H$_2$O in collision  with H. 
}
\label{3vs}
\end{figure}

Finally in Fig.~\ref{3vs} we compare the state-to-state quenching cross section from  initial states $(010)0_{00}$, 
$(100)1_{11}$, and $(001)0_{00}$ to the lowest rotational states in the ground vibrational state (000) of H$_2$O. At collision energies below $\sim 1500$ cm$^{-1}$, the transition from bending vibrational state $(010)0_{00}$, with smaller energy gap between initial and final states, displays a much larger cross section than from the initial vibrational stretching states which have similar energy levels.

Calculations of quenching from higher excited rotational and vibrational states of H$_2$O are in progress. The current results, future large-scale coupled-states (CS) approximation calculations, and machine-learning methods will be essential in the construction of a database of H$_2$O vibrational and rotational quenching rate coefficients urgently needed for astrophysical modeling.

\section{Supporting Information}

  A table comparing H$_2$O-H PES global minimum, rigid-rotor scattering results, and additional rovibrational scattering cross sections and rate coefficients of H$_2$O in collision with H.

\section{Acknowledgements}

This work was partially supported at UGA, Penn State, UNLV, and Emory by NASA grant 80NSSC22K1167;
 at the  University of New Mexico by the Department of Energy (Grant DE-SC0015997); at UNLV by NSF Grant No. PHY-2409497;  and at Penn State by NSF Grant No. PHY-1806180. This study was supported by resources and technical expertise from the UGA Georgia Advanced Computing Resource Center (GACRC).
 
{}


\begin{thebibliography}{}

\bibitem{chu69} Cheung, A. C.; Rank, D. M.; Townes, C. H.; Thornton, D. D.; Welch, W. J.
 Detection of Water in Interstellar Regions by its Microwave Radiation.
 {\it Nature} {\bf 1969}, {\it 221}, 626-628.

\bibitem{wri00}
 Wright, C. M.; van Dishoeck, E. F.; Black, J. H.; Feuchtgruber, H.; Cernicharo, J.; Gonzalez-Alfonso, E.; 
  de Graauw, Th.  ISO-SWS observations of pure rotational H$_2$O absorption lines toward Orion-IRc2.
 {\it A\&A} {\bf 2000}, {\it 358}, 689-700.                         

\bibitem{car04} Carr, J. S.; Tokunaga, A. T.; Najita, J.
 Hot H$_2$O Emission and Evidence for Turbulence in the Disk of a Young Star.
{\it ApJ} {\bf 2004}, {\it 603(1)}, 213-220.

\bibitem{nic15} Indriolo, N.; Neufeld, D. A.; DeWitt, C. N. et al.
 SOFIA/EXES Observations of Water Absorption in the Protostar AFGL 2591 at High Spectral Resolution.
{\it ApJ} {\bf 2015}, 802, L14

\bibitem{bau18} Baudry, A.; Humphreys, E. M. L.;  Herpin, F. et al.
Vibrationally excited water emission at 658 GHz from evolved stars.
{\it A\&A} {\bf 2018}, {\it 609}, A25.

\bibitem{dav17}  Neufeld, D. A.; Melnick, G. J.; Kaufman, M. J.  et al.
 SOFIA/GREAT Discovery of Terahertz Water Masers.  {\it ApJ} {\bf  2017}, {\it 843(2)}, 94.

\bibitem{asa23} Asanok, K.; Gray, M. D.; Hirota, T. et al.
 Proper Motions of Water Masers in W49 N Measured by KaVA. {\it ApJ} {\bf  2023}, {\it 943(2)}, 79.

\bibitem{van21} van Dishoeck, E. F.; Kristensen, L. E.; Mottram, J. C. et al.
Water in star-forming regions: physics and chemistry from clouds to disks as probed by Herschel spectroscopy.
 {\it A\&A} {\bf 2021}, {\it 648}, A24.  

\bibitem{bar22} Barr, A. G.; Boogert, A.; Li, J. et al.
Infrared H$_2$O Absorption in Massive Protostars at High Spectral Resolution: Full Spectral Survey
Results of AFGL 2591 and AFGL 2136.  {\it ApJ} {\bf 2022}, {\it 935}, 165.

\bibitem{li23} Li, J. L.; Boogert, A.; Barr, A. G. et al.,
High-resolution SOFIA/EXES Spectroscopy of Water Absorption Lines in the Massive Young Binary W3 IRS 5.
 {\it ApJ} {\bf 2023}, {\it 953}, 103.

\bibitem{bau23} Baudry, A.;  Wong, K. T.; Etoka, S. et al.
 ATOMIUM: Probing the inner wind of evolved O-rich stars with new, highly excited H$_2$O and OH lines.
 {\it A\&A}  {\bf 2023}, {\it 674}, A125.  

\bibitem{van23} van Dishoeck, E. F.; Grant, S.; Tabone, B. et al.
  The diverse chemistry of protoplanetary disks as revealed by JWST.
  {\it Faraday Discuss.} {\bf 2023},  {\it 245}, 52-79.
  
\bibitem{per23} Perotti, G.; Christiaens, V.; Henning, Th. et al.
 Water in the terrestrial planet-forming zone of the PDS 70 disk. {\it Nature} {\bf 2023},  {\it 620}, 516-520.

\bibitem{gas23} Gasman, D.; van Dishoeck, E. F.;  Grant, S. L. et al.
 MINDS. Abundant water and varying C/O across the disk of Sz 98 as seen by JWST/MIRI. {\it A\&A} {\bf 2023}, {\it 679}, A117.

\bibitem{ten01} Tennyson, J.; Zobov, N. F.; Williamson, R.; Polyansky, O. L.; Bernath, P. F.
  Experimental Energy Levels of the Water Molecule.
  {\it J. Phys. Chem. Ref. Data}. {\bf 2001}, {\it 30}, 735-831.

\bibitem{zha91} Zhang, Q.; Sabelli, N.; Buch, V. Potential energy surface of H$\cdots$H$_2$O.
 {\it J. Chem. Phys.} {\bf 1991}, {\it 95}, 1080-1085.

\bibitem{dag13} Dagdigian, P. J.; Alexander, M.  H.
 Exact quantum scattering calculations of transport properties for the H$_2$O-H system.
 {\it J. Chem. Phys.} {\bf 2013}, {\it 139}, 194309.

\bibitem{mcc21} McCarver, G. A.; Hinde, R. J. High accuracy ab initio potential energy surface for
  the H$_2$O-H van der Waals dimer. {\it J. Chem. Phys.} {\bf 2021}, {\it 155}, 114302.

\bibitem{cab22} Cabrera-Gonz\'{a}lez, L. D.; Denis-Alpizar, O.; P\'{a}ez-Heern\'{a}ndez, D.; Stoecklin, T.
  Quantum study of the bending relaxation of H$_2$O by collision with H.
{\it Monthly Notices of the Royal Astronomical Society}. {\bf 2022}, {\it 514,} 4426-4432.

\bibitem{gar76} Garrison, B. J.; Lester, W. A., Jr.; Miller, W. H.
  Coupled-channel study of rotational excitation of a rigid asymmetric top by atom impact:
  (H$_2$CO,He) at interstellar temperatures. {\it J. Chem. Phys.} {\bf 1976}, {\it 65(6)}, 2193-2200.

\bibitem{gre76} Green, S.  Rotational excitation of symmetric top molecules by collisions with atoms:
    Close coupling, coupled states, and effective potential calculations for NH$_3$-He.
   {\it J. Chem. Phys.}  {\bf 1976},  {\it 64(8)}, 3463-3473.

\bibitem{dan15} Daniel, F.; Faure, A.; Dagdigian, P. J.; Dubernet, M.-L.; Lique, F.;
 des Forets, G. P. Collisional excitation of water by hydrogen atoms.
{\it Monthly Notices of the Royal Astronomical Society}. {\bf 2015}, {\it 446}, 2312-2316.

\bibitem{bis93} Bissonnette, C.; Clary, D. C.
 Coupled channel calculations on rovibrational excitation of H$_2$O in collisions with H atoms
{\it Chem. Phys.} {\bf 1993}, {\it 175}, 23-36.

\bibitem{gar23} Garc\'{i}a-V\'{a}zquez, R. M.;  Faure, A.;  Stoecklin, T.
  Bending relaxation of H$_2$O by collision with para- and ortho-H$_2$.
{\it ChemPhysChem}. {\bf 2023}, {\it 25 (2),} e202300698.

\bibitem{abcd} Yang, D.; Chai, S.; Xie, D.; Guo, H.
  ABC+D: A time-independent coupled-channel quantum dynamics program for elastic and ro-vibrational
  inelastic scattering between atoms and triatomic molecules in full dimensionality.
  {\it J. Chem. Phys.} {\bf 2023}, {\it 158}, 054801.

\bibitem{yang21} Yang, D.; Xie, D.; Guo, H.
  A Time-Independent Quantum Approach to Ro-vibrationally Inelastic Scattering between Atoms and
 Triatomic Molecules. {\it J. Phys. Chem. A} {\bf 2021}, {\it 125 (31)}, 6864-6871.

\bibitem{yang22a} Yang, D.; Xie, D.; Guo, H.
   Extended coupled-states approximation for full-dimensional quantum treatments of rovibrationally
   inelastic scattering between atoms and triatomic molecules.
    {\it J. Chem. Phys.} {\bf 2022}, {\it 157 (16)}, 164111.

\bibitem{yang22b} Yang, D.; Liu, L.; Xie, D.; Guo, H.
 Full-dimensional quantum studies of vibrational energy transfer dynamics between H$_2$O and Ar:
theory assessing experiment. {\it Phys. Chem. Chem. Phys.} {\bf 2022} {\it 24(22)}, 13542-13549.

\bibitem{yang23} Yang, D.; Guo, H.; Xie, D. Recent advances in quantum theory on ro-vibrationally 
  inelastic scattering. {\it Phys. Chem. Chem. Phys.} {\bf 2023}, {\it 25}, 3577-3594.

\bibitem{bast} Brocks, G.; van der Avoird, A.; Sutcliffe, B. T.; Tennyson, J.
Quantum dynamics of non-rigid systems comprising two polyatomic fragments.
 {\it Mol. Phys.} {\bf 1983}, {\it 50}, 1025-1043.

\bibitem{man86} Manolopoulos, D. E. An improved log derivative mehtod for inelastic scattering.
  {\it J. Chem. Phys.} {\bf 1986}, {\it 85}, 6425-6429.

\bibitem{joh73} Johnson, B. R. The multichannel log-derivative method for scattering calculations.
   {\it J. Comput. Phys.} {\bf 1973}, {\it 13}, 445-449.

\bibitem{molpro} H.-J. Werner, P. J. Knowles, G. Knizia, F. R. Manby, M. {Sch\"{u}tz}, {\it et al.},
  Molpro, version 2010.1, a package of {\it ab initio} programs, 2010, see http://www.molpro.net

\bibitem{adl07} Adler, T. B.;  Knizia, G.;  Werner, H.-J.
 A simple and efficient CCSD(T)-F12 approximation. {\it J. Chem. Phys.} {\bf 2007}, {\it 127(22)}, 221106.

\bibitem{wer07} Werner, H.-J.;  Adler, T. B.;  Manby, F. R.  General orbital invariant MP2-F12 theory.
  {\it J. Chem. Phys.} {\bf 2007}, {\it 126}, 164102.

\bibitem{zhe10} Xie, Z.; Bowman, J. M.  Permutationally Invariant Polynomial Basis for Molecular Energy
  Surface Fitting via Monomial Symmetrization.
  {\it J. Chem. Theory Comput.} {\bf 2010}, {\it 6(1)}, 26-34.

\bibitem{lig00} Light, J. C.; Carrington, T. Discrete-Variable Representations and their Utilization.
  {\it Adv. Chem. Phys.} {\bf 2000}, {\it 114}, 263-310.

\end{thebibliography}
\end{document}